\documentstyle[preprint,aps]{revtex}
\tighten

\begin{document}
\draft
\preprint{\vbox{Submitted to Phys. Lett. {\bf B} \hfill FSU-SCRI-95-20 \\
                                            \null\hfill IU-NTC-95-01}}
\title{Relativistic treatment of the charged-current \\
       reaction ${}^{12}{\rm C}(\nu_{\mu},\mu^{-})$ near threshold}
\author{Hungchong Kim $^1$\footnote{Email: hung@iucf.indiana.edu},
J. Piekarewicz $^2$ \footnote{Email: jorgep@scri.fsu.edu},
C. J. Horowitz $^1$ \footnote{Email: charlie@iucf.indiana.edu}}
\address{$^1$ Nuclear Theory Center and Dept. of Physics,
Indiana University, Bloomington, Indiana 47408 \\
$^2$ Supercomputer Computations Research Institute, Florida State
University,\\
Tallahassee, Florida 32306}
\maketitle
\begin{abstract}
We compute the cross section for the charged-current reaction
${}^{12}{\rm C}(\nu_{\mu},\mu^{-})$ near threshold using a
relativistic mean-field formalism. A reduced value of the
nucleon mass in the medium --- coupled to the finite muon
mass --- leads to a 30\% reduction in the cross section
relative to a free Fermi-gas estimate. Isovector RPA
correlations, which are strongly repulsive at these
momentum transfers, reduce the cross section by an
additional 15-30\%. Hence, relativistic nuclear-structure
effects can account for the more than a factor-of-two reduction
in the cross section recently reported by the LSND
collaboration.
\end{abstract}

\narrowtext

The remarkable achievements made over the past few
years in neutrino-nucleus scattering are
challenging our theoretical understanding
of these reactions~\cite{allen90,koetk92,bodma93,alber94}.
These processes, which complement electron-scattering studies,
can broaden our understanding of
nuclear structure and constitute a fruitful ground for testing the
electroweak sector of the standard model. For example,
charged-current neutrino ($\nu,\mu$) scattering  experiments at
high momentum transfer have been used to determine the axial form
factor of the nucleon~\cite{ahren88}. This knowledge, important in its
own right, becomes essential for a reliable extraction of the
strange-quark content of the nucleon from neutral-current neutrino
($\nu,p$) scattering experiments. Indeed, the strong correlation
between the axial mass parameter, $M_{A}$, and the isoscalar
strange-form factor, $g_{A}^{s}$, complicates the extraction
of strange-quark information~\cite{garve93,horow93}. Although
the common assumption made to extract hadronic-structure
information --- that nuclear-structure effects can be
described by a Fermi-gas response --- seems justified at high
momentum transfer, we have recently reported a 10\% uncertainty
in the extracted value of $M_{A}$ due to nuclear-structure
effects~\cite{kim94}. This uncertainty is so large that
present-day experiments provide no significant evidence in favor
of a nonzero $g_{A}^{s}$.

If nothing else, nuclear-structure effects will become more
important at low-momentum transfers. Here Pauli blocking,
binding-energy corrections, and RPA correlations --- all
unimportant at high momentum transfer --- must be carefully
addressed. Moreover, low-energy neutrino experiments open
a new window into the study of the nuclear axial response.
The axial response is not accessible with electromagnetic
probes, yet it is as fundamental as the vector (longitudinal
and transverse) responses measured in electron scattering.
In addition, these reactions  represent an important diagnostic
tool for neutrino detection. Hence, the identification of novel
and interesting physics~---~such as neutrino oscillation~---
becomes strongly coupled to the study of these inclusive reactions.

Motivated by recent experiments we devote this letter
to the study of the charged-current reaction
${}^{12}{\rm C}(\nu_{\mu},\mu^{-})$ near
threshold~\cite{koetk92,alber94}. We
place particular emphasis on relativistic nuclear-structure
effects in the hope of understanding the factor-of-two
reduction in the inclusive cross section, relative to a
Fermi-gas estimate, reported recently by the Liquid
Scintillator Neutrino Detector (LSND) collaboration at
LAMPF~\cite{alber94}.

We have calculated the inclusive
charged-current cross section using a relativistic
mean-field approximation to the Walecka
model~\cite{walec74,serot86}.
The model is characterized by the presence of
strong scalar and vector mean fields which induce
large shifts in the mass ($M^{*}$) and energy of
a nucleon in the medium. We use the impulse approximation
to write the charged-current operator for a target nucleon
in terms of single-nucleon form factors parameterized
from on-shell data:
\begin{equation}
\Gamma^\mu(q)= F_1(Q^2)\gamma^\mu +
              iF_2(Q^2) \sigma^{\mu\nu} {q_\nu \over 2M} -
               G_A(Q^2)\gamma^\mu \gamma^5 +
               F_p(Q^2)q^\mu\gamma^5\  \;, \quad
               (Q^2 \equiv {\bf q}^{2}-q_{0}^{2}) \;.
\label{cur}
\end{equation}
The nuclear-structure information is contained in a large
set of nuclear response functions that are computed in
nuclear matter using a relativistic random-phase approximation
(RPA) to the Walecka model. A detailed account of the model
can be found in Ref.~\cite{kim94}

At this point it is instructive to discuss our nuclear-matter
approximation. From previous electron-scattering studies we
know that a nuclear-matter response at small momentum transfers
($|{\bf q}| < 2k_{F}$) leads to an unrealistic distribution of
quasielastic strength. Undoubtedly, our present neutrino calculations
will suffer from the same deficiencies. Indeed, the ``triangle-shaped''
responses of Fig.~\ref{figone} ---~which characterize any
nuclear-matter calculation~--- clearly display some of these
limitations. Still, nuclear-matter calculations can be quite valuable.
This value stems from the fact that present day neutrino experiments
are still unable to map out the full distribution of quasielastic
strength and, thus, are forced to combine many different measurements
into an integrated cross section. By doing this, most of the fine details
of the response are lost. It is clear, however, that with the emergence
of more sophisticated experimental facilities and techniques finite-nucleus
calculations will become imperative.

In order to understand the interplay between the different
contributions to the inclusive cross section we have plotted
in Fig.~\ref{figone} the distribution of quasielastic strength,
$d^{2}\sigma/dE_{\mu}d\Omega$, using the flux-weighted average
neutrino energy for the LSND experiment,
$E_{\nu}=180$~MeV, and a characteristic momentum transfer of
$|{\bf q}|=210$~MeV. We have modeled the ground state of ${}^{12}{\rm C}$
as a Fermi gas of nucleons, with mass $M$ or $M^{*}$, at a Fermi
momentum of $k_{F}=225$~MeV. The relativistic Fermi gas result
(impulse with M) is indicated with the solid line. For the most
part, Fermi motion is responsible for a simple redistribution of
single-particle strength. However, for the present kinematical
conditions ($|{\bf q}| < 2k_{F}$) almost 35\% of phase space is unavailable
due to Pauli blocking. This suppression of the nuclear response, in
conjunction with some kinematical constraints, generates a reduction
of almost 50\% in the cross section relative to its single-nucleon
value.

Although interesting, the main focus of the present paper is not
the Pauli suppression of the Fermi-gas cross section. Rather, it
is how additional nuclear structure effects, such as those arising
from a reduced nucleon mass and RPA correlations, induce a large
reduction in the cross section relative to its Fermi-gas value.
In this regard, it is particularly interesting to study the interplay
between the effective nucleon mass $M^{*}$ and the finite muon mass.
Thus, we start with some simple kinematical considerations. First,
the kinematical constraint, $|{\bf q}| \le (E_{\nu}+|{\bf k}^{'}_{j}|)$,
imposed at the leptonic vertex leads to a maximum allowed value for the
energy loss satisfying (for $E_\nu=180$ MeV)
\begin{equation}
  q^{0} \le E_{\nu}-\sqrt{(|{\bf q}|-E_{\nu})^{2}+m_{j}^{2}}=
  \cases{70.17~{\rm MeV}, & for muons;     \cr
         150.0~{\rm MeV}, & for electrons. \cr}
\end{equation}
Note, the subscript $j$ on the lepton momentum, $|{\bf k}^{'}_{j}|$,
represents either a muon or an electron. An additional constraint on
$q^{0}$ can be obtained from examining the hadronic vertex. In this
case the maximum allowed value corresponds to the energy transfer to
a nucleon at the Fermi surface moving in the direction of the momentum
transfer ${\bf q}$, i.e.,
\begin{equation}
  q^{0} \le \sqrt{(k_{F}+|{\bf q}|)^{2}+M^{2}}-
             \sqrt{k_{F}^{2}+M^{2}}=
  \cases{69.29~{\rm MeV}, & for $M^{*}/M=1$;    \cr
         95.67~{\rm MeV}, & for $M^{*}/M=0.68$. \cr}
\end{equation}
Note that both conditions must be simultaneously satisfied.
Among the simplest consequences of a reduced nucleon mass is a
shift in the position of the peak and an increase in the width
of the quasielastic region. Indeed, these features are clearly
visible in the impulse with $M^{*}$ calculation (dot-dashed line
in Fig.~\ref{figone}) where the position of the peak and the width of the
quasielastic region have been scaled by roughly $M/M^{*}$. Note,
however, that in the muon case the quasielastic strength is not
exhausted because of the kinematical cutoff imposed by the finite
muon mass. In particular, the interplay between $M^{*}$ and
the muon mass, together with other dynamical effects to be discussed
shortly, generate a reduction of almost 30\% in the integrated cross
section relative to the Fermi-gas value (see Table~\ref{tableone}).
For comparison, the reduction in the electron case, where the full
quasielastic strength is exhausted, becomes only 15\% [see
Fig.~\ref{figone} (b) and Table~\ref{tableone}]. This 15\%
reduction is associated
with the quenching --- due to $M^{*}$ --- of the individual nuclear
responses. A particularly interesting case is the largely untested
axial response. This response is proportional to the imaginary part
of the axial polarization defined by:
\begin{equation}
    i\Pi^{\mu 5; \nu 5}(q) =  \int{d^4p\over (2\pi)^4}\,
    {\rm Tr}[G(p+q)\,\gamma^\mu\gamma^5 \,G(p)\,
    \gamma^\nu\gamma^5] \;,
 \label{piax}
\end{equation}
where $G(p)$ is the nucleon propagator. After some simple
manipulations one can rewrite the axial polarization in terms of the
vector polarization $\Pi^{\mu\nu}$, which is probed in electron scattering,
and a leftover piece, i.e.,
\begin{equation}
    \Pi^{\mu 5; \nu 5}(q) =  \Pi^{\mu\nu}(q) +
                            g^{\mu\nu}\Pi_{A}(q) \;.
 \label{pia}
\end{equation}
Since the vector current is conserved, i.e.,
$q_{\mu}\Pi^{\mu\nu}=\Pi^{\mu\nu}q_{\nu}=0$, the violation of the
axial current is reflected by the appearance of $\Pi_{A}$. Moreover,
because this violation arises exclusively due to the presence of
a mass term, $\Pi_{A}$ must, then, be proportional to the nucleon mass.
Hence, an in-medium reduction of the nucleon mass generates a
corresponding reduction in this part of the response. In summary,
about 15\% of the quenching observed in the impulse with $M^*$ calculation
is due to a dynamical reduction of the nuclear response. The additional
15\% --- present only in the muon case --- is due to the interplay
between $M^{*}$ and the finite muon mass.

Repulsive RPA correlations are responsible for an additional,
and substantial, quenching of the cross section. We have adopted a
conventional residual interaction consisting of $\pi+\rho+g'$
contributions~\cite{kim94}. At these low momentum transfers the
longitudinal (pion-like) component of the residual interaction
has barely become attractive while the transverse (rho-like) component
remains large and repulsive. A repulsive interaction of this kind
quenches the nuclear response and shifts quasielastic strength
towards high-excitation energy (hardening). Note, the quenching and
hardening of the transverse response measured in quasielastic electron
scattering is well documented in the literature~\cite{alber82}. This
quenching and hardening of the response is clearly observed in
Fig.~\ref{figone}. For example, the position of the quasielastic peak
in the RPA with M calculation (long-dashed lines) has been shifted by more
than 30~MeV and the integrated strength reduced by almost 40\% relative to
the Fermi-gas values. The RPA reduction in the $M^{*}$ case (short dashed
line) is even more dramatic --- particularly in the muon case. Since the
repulsive RPA correlations shift strength towards high-excitation energy,
the cutoff induced by the finite muon mass now removes a substantial part
of the quasielastic strength --- reducing the integrated cross section to
only 36\% of its Fermi-gas value [see Table~\ref{tableone} under RPA(I)].
Note that this value is considerably larger (57\%) in the electron case.

In a recent publication~\cite{horpie93} we have argued in favor of a
residual interaction with a transverse component less repulsive
than the one traditionally employed in nonrelativistic calculations
and the one adopted by us, until now.  We have indicated that, because
of the in-medium reduction of the nucleon mass, a relativistic mean-field
(impulse with $M^{*}$) calculation could account for most of the features
(i.e., quenching and hardening) observed experimentally in the $(e,e')$
transverse response without the need for such a repulsive residual
interaction. Moreover, with this residual interaction we were able to
reproduce most of the spin observables measured in a recent
quasielastic $(\vec{p},\vec{n})$ experiment~\cite{mccle92,chen93}. Our
results with this residual interaction have been plotted with dots in
Fig.~\ref{figone}. As expected from a less repulsive interaction
the quenching of the cross section is not as large as before. Still, even
in this case the integrated cross section amounts, in the muon case, to
only 53\% of its Fermi-gas value [see Table~\ref{tableone} under RPA(II)].
As before, this number grows considerably, up to 71\%, in the electron case.

One can extend the calculation to other values of the momentum transfer
in the hope of mapping out the two dimensional nuclear-response surface.
This kind of analysis, which is now commonplace in electron-scattering,
is at present not available with weak probes due to the low count rate
of the reaction. Therefore, in an attempt to improve statistics
present-day experiments combine many of these measurements into an
integrated cross section. Moreover, since most of the neutrinos are
produced from the decay of pions in flight one integrates over
the experimental neutrino spectrum. In this way, many of the
detailed features that one could in principle map out are,
unfortunately, lost. Yet, some interesting properties remain.
In Fig.~\ref{figtwo} we report the angle-integrated cross
section, $d\sigma/dE_{\mu}$, folded over the LSND neutrino spectrum
$\phi(E_\nu)$, i.e.,
\begin{eqnarray}
\Bigl\langle {d\sigma \over dE_{\mu}} \Bigl\rangle=
             {\displaystyle{\int_{0}^{\infty}{d\sigma \over dE_{\mu}}} \
             \phi(E_\nu)\ dE_\nu
             \over \int_{0}^{\infty} \phi(E_\nu)\ dE_\nu } \;,
\end{eqnarray}
where the above integrations have been carried out using the full
range of the experimental spectrum, i.e., $0\le E_\nu \le 300$~MeV.
Note that the numerator has support only above the threshold value
for the ${}^{12}{\rm C}(\nu_{\mu},\mu^{-})$ reaction, or
$E_{\nu}=111.6$~MeV in the Fermi gas model.

The fact that all the models display a common spectral shape is an indication
that most of the fine details present in the double-differential cross
section have been lost. Yet, the large quenching still remains. Indeed, the
quenching in the total inclusive cross section is almost identical to the one
reported in Table~\ref{tableone}, i.e.,
\begin{equation}
 {\langle\sigma\rangle \over \langle\sigma_{\scriptscriptstyle FG}\rangle}=
  \cases{0.70(0.80), & for impulse with $M^{*}$;    \cr
         0.53(0.59), & for RPA(I)  with $M$;        \cr
         0.39(0.51), & for RPA(I)  with $M^{*}$;    \cr
         0.55(0.66), & for RPA(II) with $M^{*}$.    \cr}
\end{equation}
Here the numbers in parentheses are the appropriate ratios for electron
neutrinos. Note that the flux-averaged inclusive cross section in the Fermi-gas
limit is $\langle\sigma_{\scriptscriptstyle FG}\rangle=8.11\times 10^{-40}
{\rm cm}^{2}$ for muons and
$\langle\sigma_{\scriptscriptstyle FG}\rangle=23.13\times 10^{-40}{\rm cm}^{2}$
for electrons. Thus, the the flux-averaged inclusive cross section in the
RPA(I) with $M^{*}$ model is only
$\langle\sigma\rangle=3.16\times 10^{-40}{\rm cm}^{2}$,
or $0.39$ of its Fermi gas value, in the muon case.
Note, these cross sections were computed without Coulomb
corrections.

In summary, we have calculated the charged-current reaction
${}^{12}{\rm C}(\nu_{\mu},\mu^{-})$ near threshold in a
relativistic formalism. We have incorporated nuclear-structure
corrections arising from an in-medium reduction of the nucleon
mass and RPA correlations. The reduction of the nucleon mass
coupled to the finite muon mass prevents the quasielastic strength
from being exhausted and results in a 30\% quenching of the
inclusive cross section relative to a Fermi-gas estimate.
Repulsive RPA correlation exacerbate the effect and, depending
on the form of the residual interaction, give rise to a
reduction in the cross section anywhere from 45-60\%. Thus,
relativistic nuclear-structure effects seem to be able to
account for the factor-of-two reduction in the cross section
reported by the LSND collaboration.

The large reductions, relative to a free Fermi gas, seen in the LSND experiment
and our calculations may have important implications for atmospheric neutrino
detection.  Presently there is an anomaly in the observed ratio of atmospheric
neutrinos, $\nu_\mu/\nu_e$, which could signal neutrino
oscillations~\cite{kamiokande}.
Although the atmospheric spectrum is typically more energetic than LSND, the
average
nuclear excitation energy and momentum transfer are still modest.  Therefore we
find that our relativistic nuclear-structure effects may still be significant
for atmospheric neutrinos.  In a future work we will study how these effects
change: (1) the absolute $\mu$ or $e$ cross section, (2) the energy dependence
of
the $\mu$ quasielastic cross section and (3) the ratio of $\mu$ to $e$ cross
sections.


\acknowledgments
This research was supported by the U.S. Department of Energy
through contracts \#~DE-FG02-87ER40365, DE-FC05-85ER250000,
and DE-FG05-92ER40750.


%
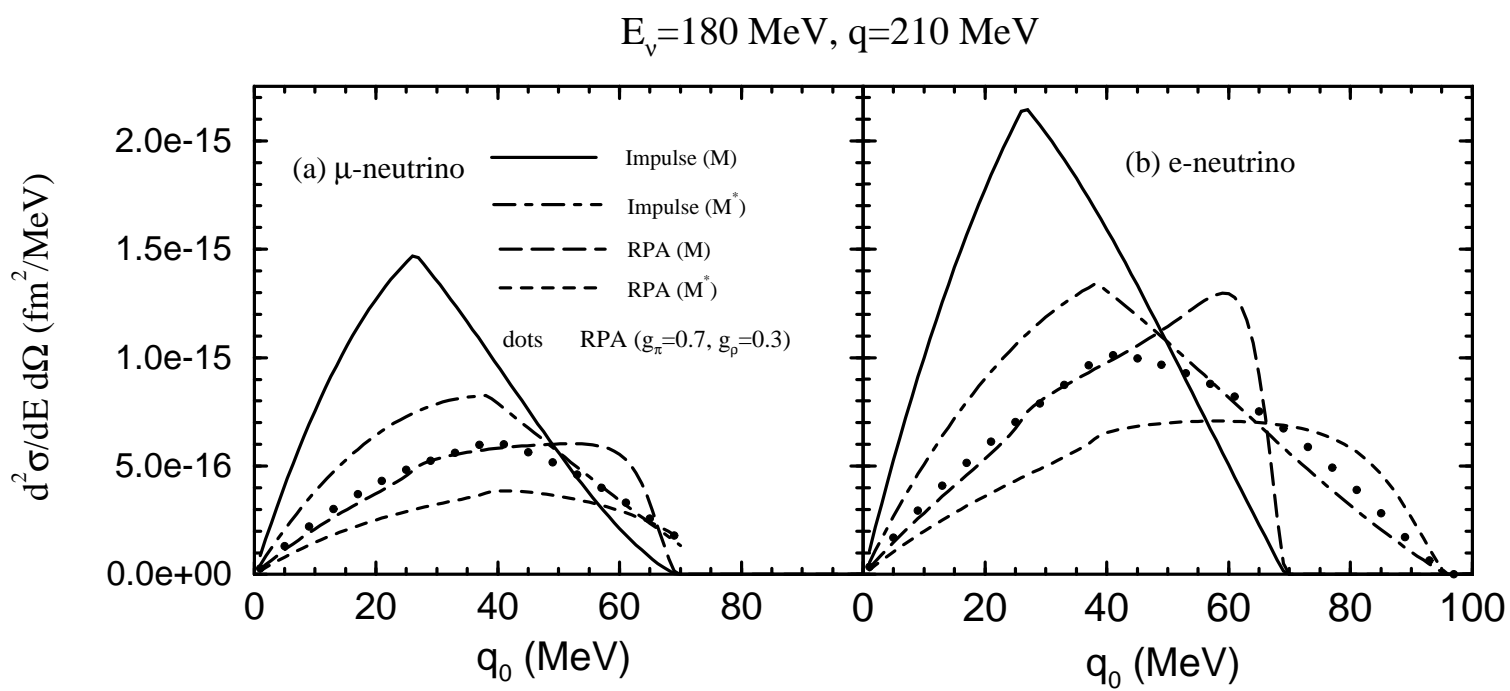
\begin{figure}
\caption{Double-differential cross section for the
         ${}^{12}{\rm C}(\nu_{\mu},\mu^{-})$ reaction
         (a) and ${}^{12}{\rm C}(\nu_{e},e^{-})$
         (b) as a function of energy loss for
         a neutrino energy of $E_{\nu}=$180~MeV,
         a momentum-transfer of $|{\bf q}|=210$~MeV, and
         a Fermi momentum of $k_{F}=225$~MeV.}
 \label{figone}
\end{figure}
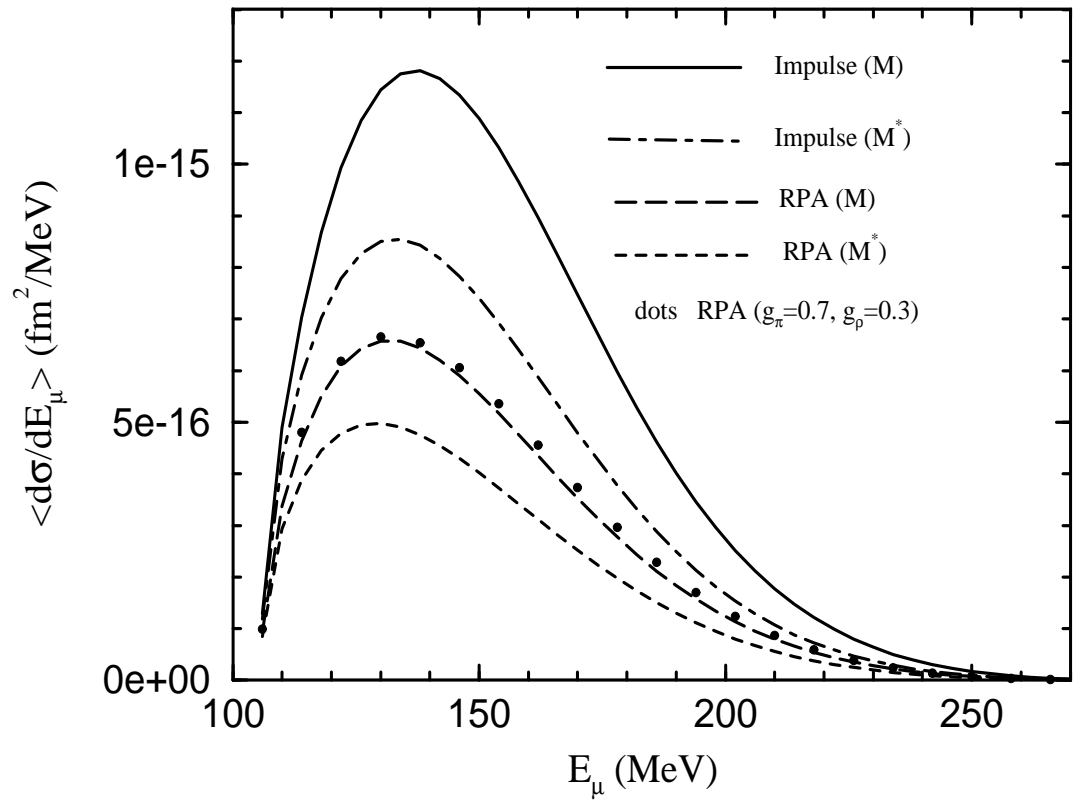
\begin{figure}
\caption{Angle-integrated cross section for the
         ${}^{12}{\rm C}(\nu_{\mu},\mu^{-})$ reaction
         folded over the LSND neutrino spectrum
         as a function of the outgoing muon energy.
         A Fermi momentum of $k_{F}=225$~MeV was used
         to simulate ${}^{12}{\rm C}$.}
 \label{figtwo}
\end{figure}
%

%
 \mediumtext
 \begin{table}
  \caption{${}^{12}{\rm C}(\nu_{\mu},\mu^{-})$ cross sections integrated over
           energy loss for an incoming neutrino energy of $E_{\nu}=180$~MeV
           and a momentum transfer of $|{\bf q}|=210$~MeV. All cross
           sections are
           reported relative to the free Fermi-gas (impulse with M) value
           using a Fermi momentum of $k_{F}=225$~MeV. (The numbers below
           are the
           areas under the curves in Fig.~\protect\ref{figone} relative to the
           area under the impulse ($M$) curve.)
           Quantities in parentheses
           are the appropriate cross sections for the
           ${}^{12}{\rm C}(\nu_{e},e^{-})$ reaction.  The area under
           the impulse
           $M$ curve of $\nu_\mu$ ($\nu_e$) is $5.213 \times 10^{-15}\
           {\rm fm}^2$ ($8.115 \times 10^{-15}\ {\rm fm}^2$). }
   \begin{tabular}{cccc}
   $M^{*}/M$ &   Impulse    &  RPA(I)       &  RPA (II)     \\
        \tableline
      1.00   & 1.00 (1.00)  & 0.56 (0.64)   &     ---       \\
      0.68   & 0.70 (0.84)  & 0.36 (0.57)   & 0.53 (0.71)   \\
   \end{tabular}
  \label{tableone}
 \end{table}
\end{document}